# A PHOTOIONIZATION MODEL FOR THE SOFT X-RAY SPECTRUM OF NGC 4151

B. K. Armentrout[1], S. B. Kraemer[1], T. J. Turner[2]

## Abstract

We present analysis of archival data from multiple *XMM-Newton* observations of the Seyfert 1 galaxy NGC 4151. Spectral data from the RGS instruments reveal several strong soft X-ray emission lines, chiefly from hydrogen-like and helium-like oxygen, nitrogen, neon and carbon. Radiative recombination continua (RRC) from oxygen and carbon are also detected. Our analysis suggests that the emission data are consistent with photoionization. Using the CLOUDY photoionization code, we found that, while a two-component, high column density model ($10^{23}$ cm$^{-2}$) with low covering factor proved adequate in reproducing all detected Lyman series lines, it proved insufficient in modeling He-like triplets observed (neon, oxygen, and nitrogen). If resonance line data were ignored, the two-component model was sufficient to match flux from intercombination and forbidden lines. However, with the inclusion of resonance line data, He-like triplets could no longer be modeled with only two components. We found that observed oxygen *G* and *R* ratios especially were anomalous in parameter space investigated. We investigated, and were forced to dismiss, the possibility that a third purely collisional component could be responsible for enhanced resonance line contributions. We succeeded in modeling the observed spectrum with the addition of a third, lower column density ($10^{20.5}$ cm$^{-2}$) component with non-zero microturbulence and high covering factor. While sufficient to reproduce observed soft X-ray flux, our model faces certain shortcomings, particularly in a less-than-ideal visual fit to the line profile.

Two of the three emission model components bear similarities to components determined by Kraemer et al. (2005) in their study of NGC 4151 absorption spectra.

Subject headings: galaxies: individual (NGC 4151) --- galaxies: Seyfert --- X-rays: individual (NGC 4151)

1. *Department of Physics, Catholic University of America, 620 Michigan Avenue NE, Washington, DC 20064*

2. *Joint Center for Astrophysics, Physics Department, University of Maryland Baltimore County, 1000 Hilltop Circle, Baltimore, MD 21250*

# 1. INTRODUCTION

NGC 4151 is the nearest bright, if not prototypical, Seyfert type 1.5 galaxy ($z = 0.003319$[1], distance = 14.20 Mpc, $H_o = 70$ km s$^{-1}$ Mpc$^{-1}$), exhibiting both narrow and broad line emission in the optical and UV spectra, and has been well-studied over all wavelengths. Historical soft X-ray observations have indicated significant complex nuclear continuum absorption (Yaqoob, Warwick & Pounds 1989, George et al. 1998), and models of X-ray absorbers subsequently predicted the presence of multiple emission lines in the spectrum below 1 keV (Netzer 1993, 1996). More recent X-ray observations indicate time-variable flux, and have further confirmed significant absorption of the nuclear continuum in the soft X-ray spectrum (Kraemer et al. 2005). In addition, contemporary kinematical models have postulated that the observational line of sight of NGC 4151 is oriented roughly halfway between the radio axis of the nuclear continuum source and the equatorial accretion disk (Crenshaw et al. 2000; Das et al. 2005). The availability of high resolution instruments onboard the *Chandra* and *XMM-Newton* X-ray observatories have yielded new insights about previously unresolved soft X-ray emission lines.

The 0.4 - 1.2 keV soft X-ray spectrum of NGC 4151 is dominated by narrow line emission and radiative recombination continua (RRC) from hydrogenic and He-like carbon, oxygen, neon, and nitrogen (Ogle et al. 2000, Schurch et al. 2004, Kraemer et al. 2005). While *XMM-Newton* Reflection Grating Spectrometer (RGS) instrumental resolution limitations place only an upper limit on emission line widths (FWHM ~ $10^3$ km s$^{-1}$), the spectrum itself is most likely created by reprocessing in the narrow emission line region (NLR). There is general agreement on the physical region of this emission, but the ionization mechanism itself continues to be a subject of debate. Photoionization and thermal collisional processes may both contribute to narrow line emission, but the extent to which any one mechanism dominates in this spectrum is not agreed upon (Ogle et al.). Fortunately, diagnostics are now available which may provide detailed information about electron temperature and number density based on RRC width (Liedahl 1999) and the relative strengths of He-like triplets (Porquet & Dubau 2000, Porter & Ferland, 2007), from which predictions can be made regarding the relative contributions of photoionization and collisional processes to emission spectra. Additional archival data available from *Chandra* and *XMM-Newton* have provided fertile new ground for analysis of the ionization processes in NGC 4151. Ogle et al. used *Chandra* data to conclude that a photoionization component must be present in NGC 4151 to account for observed narrow radiative combination continua (RRC) and high ratios of forbidden to resonance emission lines in He-like ions. Schurch et al., in their analysis of the *XMM-Newton* soft X-ray observations in 2000 upon which this paper is based, found similar evidence to suggest photoionization as a contributor to emission.

---
[1] NASA Extragalactic Database

Due to relatively low electron temperatures deduced from RRC (~ $10^4$ - $10^5$ K), it is believed that collisional ionization plays at best a minor role in excitation and ionization in NGC 4151. Can we, however, account for the observed soft X-ray spectrum entirely from photoionization? To date, no self-consistent photoionization model has been proposed to recreate the observed NGC 4151 soft X-ray spectrum. Accordingly, we analyze archival *XMM-Newton* spectra from December 2000, developing a model fit to the emission lines observed in the soft X-ray spectrum using the photoionization code CLOUDY (Ferland et al. 1998).

## 2. OBSERVATION AND DATA PROCESSING

### 2.1 Observation Data

NGC 4151 was observed by the two *XMM-Newton* RGS instruments in High Event Rate mode in five separate observations over the three-day period December 21-23, 2000. First order spectral data were reprocessed using the RGSPROC command in the *XMM-Newton* Science Analysis Software (SAS) package, version 6.1.0. The background subtraction flag was disabled during reprocessing, essentially extracting source and background spectra, allowing subsequent XSPEC fits to handle background subtraction and error propagation. Possible flare contamination was filtered by limiting data to intervals with count rates lower than 0.2 s$^{-1}$. RGS2 data have a gap over the energy range 0.51-0.62 keV due to failure of CCD chip 4 within a week of launch; RGS1 data have a gap over the energy range 0.90-1.17 keV due to in-flight failure of CCD chip 7 in September 2000.

We added data from the five observations to form a single spectral file for each instrument, using the SAS MATHPHA command, and subsequently fit the data using XSPEC v12.2.0[2]. Since significant absorption of the source continuum precludes accurate modeling of the soft X-ray power law, any fitting of the underlying power law is essentially meaningless. In order to determine accurate relative emission flux, however, we have included a twice broken power law component to determine a baseline, with break points and 506 and 580 eV. In addition, we modeled Galactic absorption with a Wisconsin absorption component, $N_H$ = 2.1 x 10$^{20}$ cm$^{-2}$ (Morrision & McCammon 1983). To model observed emission lines, we first ran a sliding Gaussian across the spectrum using the STEPPAR command, and added ZGAUSS components for each line noted. To model radiative recombination continua (RRC), we used the REDGE model component. RGS 2 data sets were constrained to those of RGS 1, with an overall CONSTANT model component added to account for instrumental variations. The CONSTANT parameter model had a value of 0.96, with a 90% confidence interval from 0.94 to 0.97. Employing $\chi$-squared statistics in our final fit, we regrouped each data set with a minimum of 20 counts per bin using the SAS GRPPHA command. Our final fit showed a reduced $\chi$-squared value of 1.55 with 1236 degrees of freedom. Table 1 shows resultant flux and widths for definitively resolved lines.

---

[2] http://heasarc.gsfc.nasa.gov/docs/xanadu/xspec/

## 2.2 Instrumental Limitations

The RGS instruments on-board *XMM-Newton* each have a wavelength accuracy of +/- 8 mÅ (XMM-Newton Users Handbook, Issue 2.3) across their bandwidth, and wavelength-dependent FWHM resolution that varies from 600 km s$^{-1}$ at 0.3543 keV (35 Å) to 1700 km s$^{-1}$ at 1.2400 keV (10 Å) in first order spectra. Since most observed emission line widths fall below this minimum resolution range in a free fit, they have not been included in our data fit. Attempts to "group-fit" lines assumed *a priori* to have similar widths (e.g., He-like resonance lines) provided no additional information.

We also note the presence of an instrumental "bump" in the RGS instrument background spectrum around a wavelength of 0.3875 keV (32 Å) (*XMM-Newton* Users Handbook). The origin of this phenomenon is not yet fully understood, but it would certainly cast doubt on definitive flux calculations for emission lines and RRC in that region, namely those of C V RRC (0.3908 keV) and C VI Ly-$\alpha$ (0.3676 keV). For this reason, although flux is clearly detected for C V RRC and C VI Ly-$\alpha$, they will not be included in our data modeling. The instrumental bump also makes definitive fitting of the power law components especially difficult.

# 3. OBSERVED EMISSION

## 3.1 Emission Line Blueshift

Previous *Chandra* observational data models (Ogle et al. 2000) and other models of the *XMM-Newton* data set we examine here (Schurch et al. 2004) note a mean blueshift in the X-ray emission spectrum, consistent with the predictions of Crenshaw et al. (2000) that suggest a biconical outflow model of the NLR, with asymmetrical physical orientation to the observer. Using archival *Hubble Space Telescope* (*HST*) Space Telescope Imaging Spectrograph (STIS) spectra, Das et al. (2005) subsequently constructed a kinematic model for NGC 4151 showing a biconical orientation that would favor measurements of bulk blueshift in emission lines in the NLR, due to far side shielding by the host galaxy. Our analysis of strong emission lines agrees with an overall blueshift in the main valued at around 200-300 km s$^{-1}$ (Table 2), although large 90% confidence intervals prevent more detailed conclusions.

## 3.2 Radiative Recombination Continua (RRC)

In observing the emitting gas surrounding an active nucleus, electron temperature can be inferred by analysis of radiative recombination continua (RRC) width ($\Delta E$) relative to edge energy ($E$) (for details, see Liedahl & Paerels 1996; Liedahl 1999). Electron temperature determined from the ratio of RRC width to edge energy can indicate either a photoionization or collisional environment. For a given ionization state, electron temperature is expected to be much higher than that for photoionization. If the ratio $\Delta E/E \ll 1$, the primary mechanism for ionization is expected to be the incident radiation field (i.e., photoionization).

Several He-like and hydrogenic RRC were conclusively detected (Table 3). Approximating temperature width in eV to $k_B T$ (Liedahl 1999; where $k_B$ is the Boltzmann constant), we calculate temperatures $\geq$ a few x $10^4$ K, noting the relatively higher temperature values for the hydrogenic C VI, O VIII, and Ne X ions, ranging from two to three times the temperature of the detected He-like ions. Since the temperature of each of the RRC are a few eV, and the edge energies are several hundred eV, the ratio of width to edge energy for each RRC is much less than unity, providing strong evidence against collisional ionization.

## 3.3 He-Like Triplets

Relative emission strength ratios of the He-like 1s2p $^1P_1 \longrightarrow$ 1s$^2$ $^1S_0$ (resonance), 1s2p $^3P_2 \longrightarrow$ 1s$^2$ $^1S_0$, 1s2p $^3P_1 \longrightarrow$ 1s$^2$ $^1S_0$ (intercombination, often blended), and 1s2s $^3S_1 \longrightarrow$ 1s$^2$ $^1S_0$ (forbidden) transitions (indicated as *r-i-f*) can indicate either a photoionization, collisional, or hybrid environment (Porquet & Dubau 2000, Bautista & Kallman 2000, Kahn et al. 2002). The resonance line arises from a purely electric dipole transition, and increases with energy above ionization threshold. Conversely, the 1s2p $^3P_1$ $\longrightarrow$ 1s$^2$ $^1S_0$ intercombination transition and forbidden line transition are both electric dipole forbidden. Since excitation collision strength is dominated by electron impact, the ratio of the sum of intercombination and forbidden intensities to resonance line intensity is inversely proportional to electron temperature. In purely photoionized plasma, resonance line intensity is relatively weak compared to the sum of the forbidden and intercombination line intensities. Electron temperature increases as the plasma becomes first a hybrid of photoionization and collisional processes, and is then finally dominated by collision. As collisional de-excitation rates increase, a decrease in radiative de-excitation occurs, first in the forbidden, and then in the intercombination line, leading to a reduction in the *(f+i)/r* ratio.

For diagnostics of He-like emission line triplets (*f,i,r*), the following are defined, where *z* represents the forbidden line flux (*f*), *x+y* the blended intercombination lines (*i*), and *w* the resonance line (*r*):

$$R = (z/(x+y)) = f/i$$

$$G = ((x+y+z)/w) = (f+i)/r$$

Three triplets were detected and modeled in XSPEC (Table 4). Porquet & Dubau (2000) note that, if the *G* ratio is less than 4, a hybrid plasma with a relatively strong resonance line is indicated, containing both photoionization/recombination and collisional processes. Our He-like line ratios show *G* values below, but close to, this threshold, suggesting at least a photoionization component. In addition, observed electron temperatures fall within the "pure photoionization" range outlined by Porquet & Dubau for each ion detected.

## 4. PHOTOIONIZATION MODELING

### 4.1 CLOUDY PARAMETERS AND CONSTRAINTS

While previous models based on emissivity tables have been proposed to reflect the soft X-ray emission spectrum of NGC 4151 (Schurch et al., 2004), no self-consistent photoionization model has been employed to date. Accordingly, we have made use of the CLOUDY photoionization modeling code[3] (Ferland et al. 1998) to recreate NGC 4151 soft X-ray emission.

---

[3]http://www.nublado.org/

Using CLOUDY version C07.02.00, we attempted to produce the observed spectrum, assuming solar elemental abundances (e.g., Grevesse & Anders 1989). The log values of elemental abundances, relative to H by number, are as follows. He: -1.00; C: -3.47; N: -3.92; O: -3.17; Ne: -3.96; Na: -5.69; Mg: -4.48; Al: -5.53; Si: -4.51; S: -4.82; Ar: -5.40; Ca: -5.64; Fe: -4.40; and Ni: -5.75. Varied parameters include the ionization parameter ($U$), which corresponds to the dimensionless ratio of hydrogen-ionizing photon to total-hydrogen densities, hydrogen column density ($N_H$, in cm$^{-2}$), hydrogen number density ($n_H$, in cm$^{-3}$), microturbulent velocity in km s$^{-1}$, and a dimensionless geometric covering factor ($C_f$). The ionization parameter is related to flux, hydrogen number density, and radius to the model slab's inner face by the formula:

$$U = \frac{Q}{4\pi r^2 c n_H}$$

Here, $Q$ represents number of hydrogen-ionizing photons emitted per second by a central source of luminosity $L_v$:

$$Q = \int_{13.6eV}^{\infty} \left(\frac{L_v}{hv}\right) dv$$

In constructing a photoionization model to reproduce observed emission data, we immediately observed that a *single* model proved insufficient to recreate the He-like and hydrogenic neon, oxygen, and nitrogen emission observed . One model was adequate to reproduce lines with relatively low ionization potentials, most notably the strong O VII and N VI forbidden lines (0.5611 and 0.4199 keV, respectively). However, this model proved insufficient in recreating higher ionization potential emission lines, including the O VIII and Ne X Ly-α lines (0.6537 and 1.0219 keV, respectively). Relative ionic abundances are highly sensitive to the ionization parameter; a comparatively low ionization parameter of unity, for instance, favors He-like states of the observed elements. As the ionization parameter is increased to 10, completely ionized states dominate, followed by hydrogenic and He-like states. If relative ionic abundances favor He-like emission, hydrogenic emission is negligible, and vice versa, and yet, significant lines were noted for both states. Therefore, two components were required, one with a comparatively lower ionization parameter ("Medium"), favoring He-like ionic states, and a second, with a sufficiently higher ionization parameter ("High"), in which hydrogenic ionic states dominate.

Assuming isotropic emission, we were able to directly compare predicted luminosity at the source with observed luminosity. By integrating predicted CLOUDY model energy flux over $4\pi$ sr at the calculated inner slab face radius, we were constrained to a global geometric covering factor ($C_f$) between zero and unity required to produce the observed emission luminosity. We note that, if the biconical structure physically modeled by Das et al. (2005) and Crenshaw et al. (2005) is accurate (see §3.2), with a near ionization cone inclined toward the observer, and a far ionization cone located behind the host galaxy, screening may reduce observed emission line flux by as much as one-half.

Since XSPEC fits of the soft X-ray power law often prove unreliable due to significant continuum absorption (§2.2), we have adopted the broken power law for spectral energy distribution (SED) of the form $L_\nu \propto \nu^\alpha$ proposed by Kraemer et al. (2005) as follows: $\alpha = -1.0$ for energies <13.6 eV, $\alpha = -1.3$ over the range 13.6 eV < h$\nu$ < 0.5 keV, and $\alpha = -0.5$ above 0.5 keV. With normalization, the broken power law described equates to a $Q$ value of 1.1e53 s$^{-1}$.

Total luminosity observed required a significant column density in order to ensure a source covering factor less than unity. To determine minimum required hydrogen column density, we selected the strong O VII forbidden line (0.5611 keV) as our luminosity benchmark. We found that, given the observed luminosity in this line, a minimum column density value of $10^{21}$ cm$^{-2}$ in our grid was required so that the source covering factor remained at or below unity, independent of other model parameters. In order to adequately represent the observed spectrum of neon, oxygen, and nitrogen, we found $10^{23}$ cm$^{-2}$ to be an optimal column density for our two-component model.

We then examined possible parameter combinations that would faithfully recreate observed He-like *G* and *R* ratios. We were able to model the observed Ne IX and N VI *G* and *R* ratios within a wide range of parameter space investigated; however, the O VII *G* and *R* ratio values presented particular difficulties. The extremely high O VII *R* ratio occupies only a small portion of investigated parameter space, and could only be modeled with a column density of $10^{22}$ cm$^{-2}$ or higher. However, such a high O VII *R* ratio created high *G* ratios for all three triplets, corresponding to low resonance line luminosity. We would expect similar fitting difficulty for the neon and nitrogen triplets, since the O VII ionization potential lies between that of N VI and Ne IX; however, this was not the case. We hypothesized that the particular case of extremely high O VII *R* ratios was due to the fact that its assumed elemental abundance is roughly five times that of the other two noted elements. Accordingly, we ran test models over column densities from $10^{21.5}$ to $10^{23.5}$ cm$^{-2}$, matching oxygen elemental abundance to that of nitrogen. We found that O VII *R* ratio behavior matched that of neon and nitrogen with roughly equal elemental abundances.

Porter & Ferland (2007) show that elevated *R* ratios in O VII can result from high column densities. Porter & Ferland demonstrate that, in such an environment, the optical depth of the intercombination line increases, effectively increasing the *R* ratio. Also, they show that low O VII *G* ratios can be achieved at low column densities through enhancement of the resonance line by continuum pumping. However, we found that the required low column density for such resonance line enhancement was insufficient to produce observed resonance emission line flux. We were therefore forced to look elsewhere for a mechanism to boost resonance emission at high column densities. Figure 1 shows Ne IX, N VI and O VII *R* ratios produced by CLOUDY with our assumed underlying continuum as a function of column density and ionization parameter for microturbulence values of 0 and 200 km s$^{-1}$.

Since our composite two-component model underpredicted resonances lines for all three species detected, we were led to incorporate a third component that would enhance resonance emission relative to other emission features. Ogle et al. (2000) proposed that a hybrid collisional/photoionization environment may be responsible for observed X-ray emission in NGC 4151. To pursue this scenario, we considered the possibility that enhanced resonance line emission could be created from a purely collisional component. To recreate purely collisional model emission, we constructed a non-photoionization component at a temperature sufficient to favor the He-like oxygen state (the potential required to ionize oxygen from O VI to O VII is 138 eV, or 1.58e6 K). We created a collisional component fixed at this temperature at column densities of $10^{20}$ through $10^{23}$ cm$^{-2}$ and number density of $10^6$ cm$^{-3}$. We subsequently found that the fit to observed emission was poor; Ne IX resonance line emission was grossly underpredicted, as were all three Ly-β lines observed. Further, temperatures inferred from radiative recombination continua are well below those required for this component (Table 5). We were ultimately forced to abandon a hybrid model as a viable predictor of observed RRC and line emission.

While forbidden and intercombination emission lines in observed He-like triplets remain relatively insensitive to microturbulence up to tested limits, resonance lines are highly sensitive to this parameter. In their analysis of *Chandra* observations of Markarian 3, Sako et al. (2000) suggested that the relative strength of resonance lines in He-like triplets could be increased due to photoexcitation by the AGN continuum. Although temperature-dependent strengthening of these lines is a contributing factor, microturbulence may add additional intensification of resonance lines in a plasma in photoionization equilibrium (PIE). We varied the CLOUDY *TURBULENCE* parameter to create a *third* model component ("Low") that would successfully enhance resonance line emission, correspondingly depressing *G* ratios to acceptable values. Figure 2 shows Ne IX, N VI and O VII *R* ratios produced by CLOUDY with our assumed underlying continuum as a function of column density and ionization parameter for microturbulence values of 0 and 200 km s$^{-1}$. We found that a comparatively low ionization parameter, low column density, high covering factor component was required, with acceptable microturbulence values of 200-800 km s$^{-1}$, anti-correlated with required component covering factor. This anti-correlation can be attributed to narrowing of the distribution of photoexciting photons available as microturbulence is decreased, necessitating an increased covering factor to provide adequate resonance line luminosity. However, since this model component contributes significantly to OVII He-triplet resonance line luminosity, we would expect that, for microturbulence values at or above roughly 200 km s$^{-1}$, noticeable spreading in the line's gaussian tails would begin to be apparent. We noticed no such phenomenon in the observed resonance lines, and we have limited microturbulence values in the Low component to 200 km s$^{-1}$. Figure 3 shows XSPEC model fits to the O VII 0.5740 keV He-like triplet resonance line for a single gaussian free fit, and for two-gaussian composites with a narrow component and wide component fixed at microturbulent velocities of 200 and 400 km s$^{-1}$. The peak is underpredicted in all models tested, including the single gaussian free fit, and any increase in line width from microturbulence exacerbates this phenomenon. We also found that the addition of microturbulence in the "High" component provides a better fit to the high-ionization potential Lyman series lines.

The issue of the O VII resonance line visual fit presents the single most significant challenge to our model. Ideally, flux ratios are best fit with a high microturbulence component, since the O VII *R* ratio becomes underpredicted by lower microturbulence values due to overprediction of the intercombination line. However, it is evident from the visual gaussian fit that even a 200 km s$^{-1}$ component diverges from the ideal. Whether due to a non-photoionization component, or perhaps an ill-defined line spread profile at this wavelength, the data quality prevents more detailed conclusions at this time.

Within the parameter space of our grid, we found hydrogen number density to be an unconstrained parameter as long as it remained below critical density values for each of the species of interest (approximately $10^9$ - $10^{10}$ cm$^{-3}$; for details, see Porquet & Dubau 2000, Porter & Ferland 2007). However, once hydrogen column density is constrained, number density becomes limited by the requirement that the model slab thickness ($\Delta R = N_H/n_H$) divided by the radius to the cloud face be less than unity (the "thin shell" model requirement assuming plane-parallel geometry; see Kraemer et al. 2005). In addition, the He-like triplet $R$ ratios are highly sensitive to hydrogen number density. (For further theoretical details, see Porter & Ferland).

Within the individual constraints of each model component, we optimized the combined multi-component model to provide the best fit to observed spectra. Results are found in Table 5. Figure 4 shows the contributions of each of the model components to total predicted luminosity for each emission line of interest. Table 6 shows a comparison of predicted composite model luminosity to observed luminosity, and Figure 5 shows the ratio of model to observed luminosities for the emission lines of interest. We have also included ratios of pertinent model to observed luminosities for *Chandra* data provided in Ogle et al. (2000), in the same figure.

## 4.2 Model Electron Temperatures, Radii, Gas Pressures

The three-component CLOUDY composite model predicts electron temperatures of 5.2e4 K (Low), 6.0e4 K (Medium), and 1.1e6 K (High). These values are well within the range consistent with photoionization, and generally agree with measured temperatures from RRC widths found in Table 3.

Calculated inner slab face radius is a function of the total number of hydrogen-ionizing photons emitted by the central object ($Q$, in s$^{-1}$), ionization parameter ($U$), and total hydrogen density ($n_H$, in cm$^{-3}$) (cf. §4.1). Assuming a $Q$ value of 1.1e53 s$^{-1}$, the three model constituents lead to calculated inner slab face radii of $R_{LOW}$ = 9.6e17 cm (0.31 pc), $R_{MEDIUM}$ = 1.7e17 cm (5.5e-2 pc), and $R_{HIGH}$ = 1.2e16 cm (3.9e-3 pc).

The physical thickness of the modeled slab ($\Delta R$) can be determined by the ratio of total hydrogen column density ($N_H$, in cm$^{-2}$) to total hydrogen number density ($n_H$, in cm$^{-3}$). From CLOUDY ideal model parameters, we calculate slab thicknesses for the three components of $\Delta R_{LOW}$ = 3.2e14 cm (1.0e-4 pc), $\Delta R_{MEDIUM}$ = 1.0e16 cm (3.2e-3 pc), and $\Delta R_{HIGH}$ = 1.0e15 cm (3.2e-4 pc). The respective ratios of physical slab thickness to radial distance to inner slab face are thus 3.3e-4 (Low), 5.9e-2 (Medium), and 8.3e-2 (High).

Since the ratio of physical slab thickness to radial distance to inner slab face is ultimately inversely proportional to the square root of $n_H$, we can establish lower limits on $n_H$ by setting this radial ratio to unity, keeping all other parameters fixed. This requires a minimum log $n_H$ value in the Low component of -0.965, a minimum Medium component log value of 4.53, and a minimum High component log $n_H$ value of 5.83. Keeping all other model parameters constant, these values predict maximum radii to inner slab faces of $R_{LOW}$ = 2.9e21 cm (940 pc), $R_{MEDIUM}$ = 2.9e18 cm (0.94 pc), and $R_{HIGH}$ = 1.5e17 cm (0.049 pc), with corresponding maximum slab thicknesses.

Our model predicts gas pressures of 1.7e-5 dynes cm$^{-2}$ (Low), 4.7e-5 dynes cm$^{-2}$ (Medium), and 3.3e-2 dynes cm$^{-2}$ (High). This implies that the Low and Medium model components may be collocated, and are in pressure equilibrium.

## 4.3 Predicted UV Emission

Table 7 shows CLOUDY predicted UV luminosities compared against Hopkins Ultraviolet Telescope (HUT) observations on December 8, 1990 (Kriss et al. 1992). With the exception of the N V 1240Å line, our X-ray photoionization model components are not a significant contributor to UV emission. We note that large error bars in the HUT observed N V line may still be consistent with this scenario.

The blended O VI 1032/1038Å line luminosity predicted by our model represents roughly 15 percent of that observed by Kriss et al. However, the observed emission includes significant flux from the broad emission line region (BLR). For a more rigorous examination, we compared the predicted model luminosity to *FUSE* observation data, in which a low state broad-line subtracted O VI 1032/1038Å flux of 2e-12 ergs cm$^{-2}$ s$^{-1}$ was calculated, corresponding to a luminosity of 4.83e40 ergs s$^{-1}$ (D. M. Crenshaw, private communication). Our model-predicted luminosity represents over 150% of the observed broad-line subtracted value. Even accounting for a possible observational deficit of a factor of two due to far-side galaxy emission (cf. §4.1), we would be forced to conclude that our Low and Middle model component regions, which produce nearly all modeled emission for this line blend, are primarily responsible for the vast majority of all observed O VI 1032Å and 1038Å emission.

Interestingly, similarities can be seen between our emission model components and the X-ray absorption components determined by Kraemer et al. (2005) in their analysis of simultaneous UV and X-ray spectra from NGC 4151. They note an "X-High" component (log $U$ = 1.05, log $N_H$ = 22.5) and a "D+Ea" component (log $U$ = -0.27, log $N_H$ = 22.46) as chiefly responsible for X-ray absorption. These components are generally correlated to our Medium and High components, which have similar column densities, and a lower and higher ionization parameter, respectively.

## 5. CONCLUSIONS

Using the CLOUDY photoionization code, we were able to successfully reproduce the soft X-ray spectrum of NGC 4151. Three components, with disparate ionization parameters, were required to model the entire data set. Two components (labeled "Medium" and "High") were sufficient to produce hydrogenic neon, oxygen, and nitrogen emission, as well as forbidden and intercombination spectra from He-like triplets of these three elements. However, resonance lines were underpredicted with these two components; in particular, the O VII He-like triplet exhibited anomalous $G$ and $R$ ratios that proved difficult to recreate in model parameter space. We attribute this phenomenon to the large oxygen elemental abundance relative to neon and nitrogen. The data required a large column depth component to account for observed triplet spectra. To adequately model $G$ and $R$ ratios for the observed elements, a third model component (labeled "Low") was required.

Although electron temperatures deduced from RRC widths of $10^4$ - $10^5$ K were insufficient to suggest a collisional component, we attempted to account for the resonance line excess with a fixed temperature, non-photoionization component of a few times $10^6$ K, which would favor the O VII ionization state. We found that even this artificially conceived model was a poor fit to observed resonance line spectra. Subsequently, we produced a superior fit to resonance line flux by introducing a third photoionization component that included microturbulence. Specifically, we found satisfactory fits to all significantly detected He-like triplets and Lyman series lines by a composite three component model (see Table 5). Although we were able to model emission line flux with the composite model, we note a less than ideal *visual* fit to the O VII resonance line profile (the strongest triplet resonance line observed). The model gaussian profile is less peaked than the observed line, with somewhat broader wings. Whether the poor fit is due to a non-photoionzation component, a poorly-defined instrumental line spread function, or some other cause, is undetermined at this time. We hope to examine the issue more closely with future *XMM-Newton* and *Chandra* observation data.

Hydrogen number density was unconstrained up to critical values of approximately $10^9$-$10^{10}$ cm$^{-3}$; however, the "thin spherical shell" plane-parallel model geometry assumption placed additional constraints on this parameter, subject to optimized ionization parameter and hydrogen column density values. Since both radial distance to the inner slab face and slab thickness are related to the aforementioned parameters, we were able to deduce these values from our optimized values. The optimized radial distance to the Low component inner slab face is 9.6e17 cm (0.31 pc), with slab thickness of 3.2e14 cm (1.0e-4 pc). The optimized radial distance to the Medium component inner slab face is 1.7e17 cm (5.5e-2 pc), with slab thickness of 1.0e16 cm (3.2e-3 pc). The optimized radial distance to the High component inner slab face is 1.2e16 cm (3.9e-3 pc), with slab thickness of 1.0e15 cm (3.2e-4 pc). Since the hydrogen number density is a relatively unconstrained parameter, we were able to determine maximum radius values by setting the ratio of slab thickness to distance to inner slab face to its maximum value of unity. With this assumption, the maximum radial distance to the Low component inner slab face and corresponding slab thickness is 2.9e21 cm (940 pc); the maximum radial distance to the Medium component inner slab face is 2.9e18 cm (0.94 pc); the maximum radial distance to the High component inner slab face is 1.5e17 cm (0.049 pc). Assuming that the BLR extends to roughly four light days (0.003 pc, Clavel 1991), our radial predictions place all model components outside the BLR. We would thus expect little or no short time period variability in this emission profile.

We gratefully acknowledge the assistance of D. M Crenshaw for *FUSE* O VI narrow-line emission data, R.L. Porter and G.J. Ferland for their assistance with CLOUDY modeling and O VII *R* ratio data interpretation, and the *NASA*/GSFC *XMM-Newton* Guest Observer Facility (GOF) and the European Space Agency *XMM-Newton* Helpdesk in processing archival data. Archival *XMM-Newton* data was obtained through the HEASARC on-line service, provided by NASA/GSFC. This research has made use of the NASA/IPAC Extragalactic database (NED), operated by the Jet Propulsion Laboratory, California Institute of Technology, under contract with NASA.

# TABLES

Table 1: NGC 4151 Spectral Fit

| Emission Line ID | Redshift Corrected Energy (keV)[1] | Rest Energy (keV) | Photon Flux ($10^{-5}$ s$^{-1}$ cm$^{-2}$)[1] | Sigma (km s$^{-1}$)[1] | |
|---|---|---|---|---|---|
| N VI (f) | 0.4202 ⁠0.4201 0.4203 | 0.4199 | 10.674 ⁠9.839 11.510 | 182 | 101 257 |
| N VI (i) | 0.4266 ⁠0.4265 0.4269 | 0.4264 | 1.390 ⁠0.950 1.855 | | |
| N VI (r) | 0.4310 ⁠0.4309 0.4311 | 0.4307 | 5.547 ⁠4.861 6.249 | | |
| C VI (β) | 0.4360 ⁠0.4358 0.4361 | 0.4356 | 2.505 ⁠2.253 3.005 | | |
| N VII (α) | 0.5008 ⁠0.5007 0.5009 | 0.5004 | 8.506 ⁠7.730 9.305 | 325 | 175 466 |
| O VII (f) | 0.5617 ⁠0.5616 0.5617 | 0.5611 | 44.947 ⁠43.129 46.765 | | |
| O VII (i) | 0.5691 ⁠0.5689 0.5692 | 0.5687 | 6.960 ⁠6.075 7.850 | | |
| O VII (r) | 0.5745 ⁠0.5742 0.5745 | 0.5740 | 14.250 ⁠13.201 15.368 | | |
| N VII (β) | 0.5937 ⁠0.5934 0.5940 | 0.5930 | 1.560 ⁠1.110 2.009 | | |
| O VIII (α) | 0.6541 ⁠0.6540 0.6542 | 0.6537 | 15.900 ⁠15.155 16.645 | 172 | 8 249 |
| O VIII (β) | 0.7749 ⁠0.7746 0.7752 | 0.7747 | 2.837 ⁠2.501 3.173 | 403 | 190 573 |
| Ne IX (f) | 0.9064 ⁠0.9060 0.9067 | 0.9052 | 5.612 ⁠5.022 6.203 | | |
| Ne IX (i) | 0.9172 ⁠0.9165 0.9179 | 0.9151 | 2.314 ⁠1.763 2.796 | | |
| Ne IX (r) | 0.9229 ⁠0.9222 0.9236 | 0.9221 | 2.552 ⁠1.953 3.118 | | |
| Ne X (α) | 1.0224 ⁠1.0212 1.0232 | 1.0219 | 2.845 ⁠2.360 3.319 | | |

*1. 90% confidence intervals.*

Table 2: Emission Line Blueshift

| Line ID | Observed Host Galaxy Redshift-Corrected Line Energy (keV)[1] | | Rest Energy (keV) | Velocity (km s$^{-1}$)[1] | |
| --- | --- | --- | --- | --- | --- |
| O VII (f) | 0.5617 | 0.5616 | 0.5611 | -321 | -267 |
| | | 0.5617 | | | -321 |
| Ne IX (f) | 0.9064 | 0.9060 | 0.9052 | -398 | -265 |
| | | 0.9067 | | | -497 |
| N VI (f) | 0.4202 | 0.4201 | 0.4199 | -214 | -143 |
| | | 0.4203 | | | -286 |
| C VI (β) | 0.4360 | 0.4358 | 0.4356 | -275 | -138 |
| | | 0.4361 | | | -344 |
| N VII (α) | 0.5008 | 0.5007 | 0.5004 | -240 | -180 |
| | | 0.5009 | | | -300 |
| N VII (β) | 0.5937 | 0.5934 | 0.5930 | -354 | -202 |
| | | 0.5940 | | | -506 |
| O VIII (α) | 0.6541 | 0.6540 | 0.6537 | -184 | -138 |
| | | 0.6542 | | | -229 |
| O VIII (β) | 0.7749 | 0.7746 | 0.7747 | -77 | 39 |
| | | 0.7752 | | | -194 |

*1. 90% Confidence Intervals*

Table 3: RRC Temperature

| Ion | Temperature (eV)[1] | | Temperature ($10^4$ K)[1] | |
|---|---|---|---|---|
| C VI | 5.3 | 5.2 – 6.0 | 6.1 | 6.0 – 7.0 |
| N VI | 0.4 | - – 1.9 | 0.5 | - – 2.2 |
| O VII | 2.6 | 2.3 – 3.1 | 3.1 | 2.6 – 3.6 |
| O VIII | 9.9 | 7.7 – 12.3 | 11.5 | 8.9 – 14.3 |
| Ne IX | 4.8 | - – 10.4 | 5.6 | - – 12.0 |
| Ne X | 9.2 | 3.7 – 14.8 | 10.7 | 4.3 – 17.1 |

*1. 90% confidence intervals.*

Table 4: He-Like Triplet Diagnostics

| Triplet | f (z)[1] | | i (x+y)[1] | | r (w)[1] | | G (f+i)/r | | R f/i | |
|---|---|---|---|---|---|---|---|---|---|---|
| O VII | 40.40 | 38.77[2] – 42.03 | 6.34 | 5.53 – 7.15 | 13.10 | 12.14 – 14.13 | 3.6 | 3.1 – 4.1 | 6.4 | 5.4 – 7.6 |
| N VI | 7.18 | 6.62 – 7.74 | 0.95 | 0.65 – 1.27 | 3.83 | 3.35 – 4.31 | 2.1 | 1.7 – 2.7 | 7.6 | 5.2 – 11.9 |
| Ne IX | 8.14 | 7.28 – 9.00 | 3.39 | 2.58 – 4.10 | 3.77 | 2.89 – 4.61 | 3.1 | 2.1 – 4.5 | 2.4 | 1.8 – 3.5 |

1. ($10^{-14}$ ergs s$^{-1}$ cm$^{-2}$)
2. 90% confidence intervals

Table 5: CLOUDY Model Parameters

|        | Log U | Log $N_H$ | Log $n_H$ | Microturbulence (km s$^{-1}$) | $C_f$ |
|--------|-------|-----------|-----------|-------------------------------|-------|
| Low    | -0.5  | 20.5      | 6         | 200                           | 0.700 |
| Medium | 0.0   | 23.0      | 7         | 0                             | 0.035 |
| High   | 1.3   | 23.0      | 8         | 200                           | 0.110 |

Table 6: Model and Observed Luminosity Data ($10^{38}$ ergs s$^{-1}$)

| | Model Luminosity | Observed Luminosity[1] | | Model Error[1] | |
|---|---|---|---|---|---|
| N VI (f) | 16.09 | 17.33 | 15.97 | -1.24 | 0.12 |
| | | | 18.68 | | -2.59 |
| N VI (i) | 2.52 | 2.29 | 1.57 | 0.23 | 0.95 |
| | | | 3.06 | | -0.54 |
| N VI (r) | 12.33 | 9.24 | 8.09 | 3.09 | 4.24 |
| | | | 10.40 | | 1.93 |
| C VI (β) | 6.14 | 4.22 | 3.80 | 1.92 | 2.34 |
| | | | 5.06 | | 1.08 |
| N VII (α) | 11.84 | 16.46 | 14.95 | -4.62 | -3.11 |
| | | | 18.00 | | -6.16 |
| O VII (f) | 98.29 | 97.50 | 93.55 | 0.79 | 4.74 |
| | | | 101.44 | | -3.15 |
| O VII (i) | 18.61 | 15.30 | 13.36 | 3.31 | 5.25 |
| | | | 17.26 | | 1.35 |
| O VII (r) | 29.60 | 31.62 | 29.29 | -2.02 | 0.31 |
| | | | 34.10 | | -4.50 |
| N VII (β) | 2.65 | 3.58 | 2.54 | -0.93 | 0.11 |
| | | | 4.61 | | -1.96 |
| O VIII (α) | 42.78 | 40.18 | 38.30 | 2.60 | 4.48 |
| | | | 42.06 | | 0.72 |
| O VIII (β) | 6.50 | 8.50 | 7.49 | -2.00 | -0.99 |
| | | | 9.50 | | -3.00 |
| Ne IX (f) | 15.28 | 19.64 | 17.58 | -4.36 | -2.30 |
| | | | 21.71 | | -6.43 |
| Ne IX (i) | 4.89 | 8.19 | 6.24 | -3.30 | -1.35 |
| | | | 9.89 | | -5.00 |
| Ne IX (r) | 10.49 | 9.10 | 6.96 | 1.39 | 3.53 |
| | | | 11.12 | | -0.63 |
| Ne X (α) | 11.05 | 11.24 | 9.32 | -0.19 | 1.73 |
| | | | 13.11 | | -2.06 |

*1. 90% confidence intervals.*

Table 7: CLOUDY UV luminosity compared to 1990 *HUT* observations ($10^{38}$ ergs s$^{-1}$)

| | CLOUDY Predicted Luminosity | 1990 HUT Observed Luminosity [1] | |
|---|---|---|---|
| He II (1085Å) | 6.3 | 41 | 17 – 65 |
| O VI (1032/1038) | 738 | 4900 | 4500 – 5200 |
| N V (1240) | 39 | 7.2 | 0 – 320 |
| N IV (1487) | 5.4 | 170 | 100 – 240 |
| C IV (1549) | 90 | 510 | 320 – 700 |
| He II (1641) | 10 | 430 | 360 – 510 |

*1. 90% confidence intervals*

# FIGURE CAPTIONS

Figure 1: *R* Ratio Contour Plots

*Figures 1a-1f show contour plots for the Ne IX, N VI, and O VII He-like triplet R ratios as a function of column density (n(H)) and ionization parameter (U) for microturbulence values of 0 and 200 km s$^{-1}$.*

Figure 2: *G* Ratio Contour Plots

*Figures 2a-2f show contour plots for the Ne IX, N VI, and O VII He-like triplet G ratios as a function of column density (N(H)) and ionization parameter (U) for microturbulence values of 0 and 200 km s$^{-1}$.*

Figure 3: O VII resonance line fit to model gaussians.

*In Figure 3a, the line is freely fit with a single gaussian with no parameter constraints. In Figure 3b, the line is fit to two gaussians. The first gaussian, representing 90% of total flux, is fixed at a sigma of 200 km s$^{-1}$. The second gaussian, representing the remainder of total flux, is fixed at a sigma of 0.5 km s$^{-1}$. Figure 3c: The same as 3b, except the dominant gaussian is fixed at 400 km s$^{-1}$.*

Figure 4: CLOUDY Three-Component Model Luminosity

*Respective contributions to overall luminosity from Low, Medium, and High model components.*

Figure 5: Relative Luminosity (CLOUDY/Observed).

*XMM-Newton data are represented by diamonds (with error bars), Chandra data (Ogle et al. 2000) by squares.*

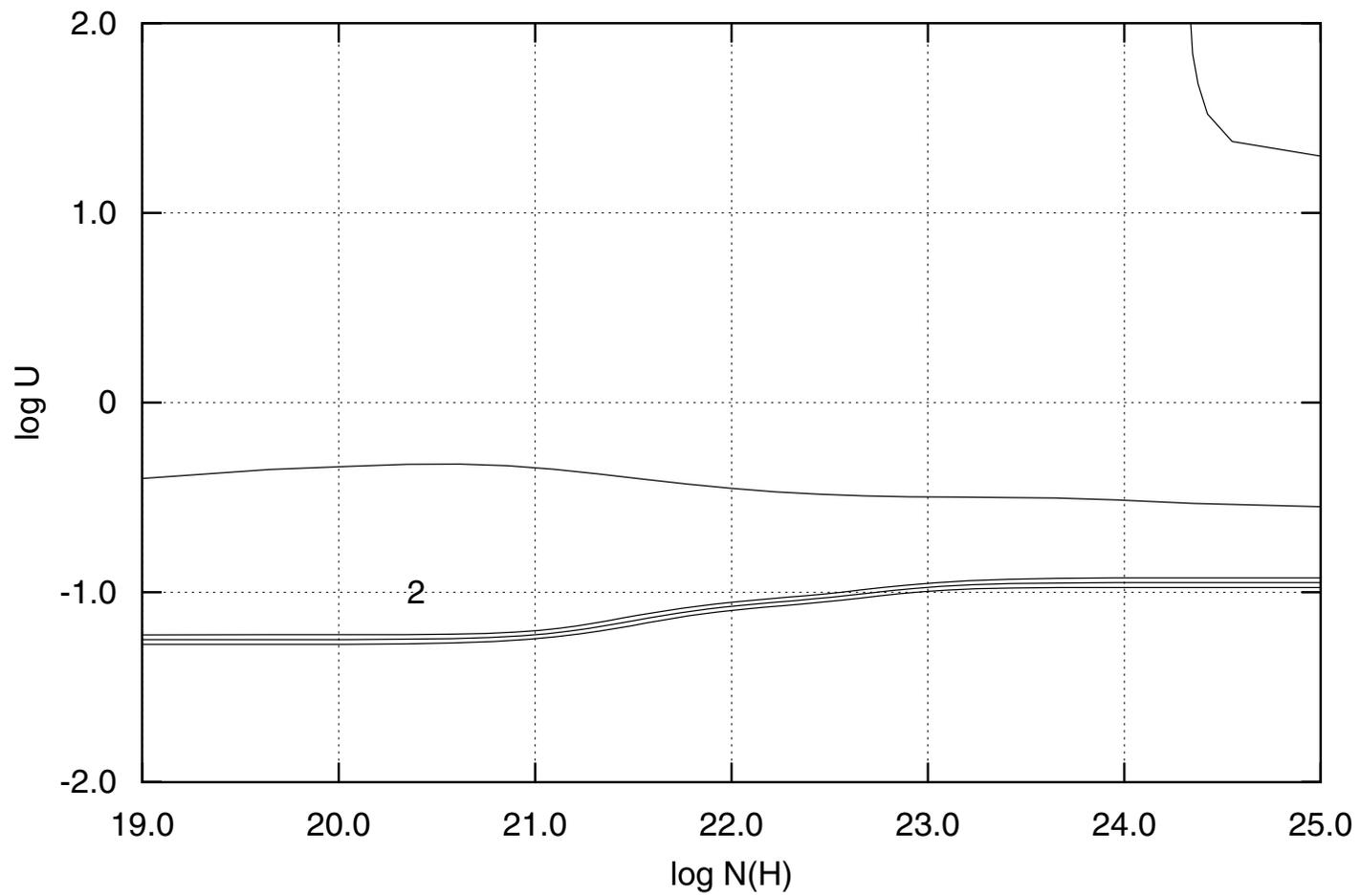

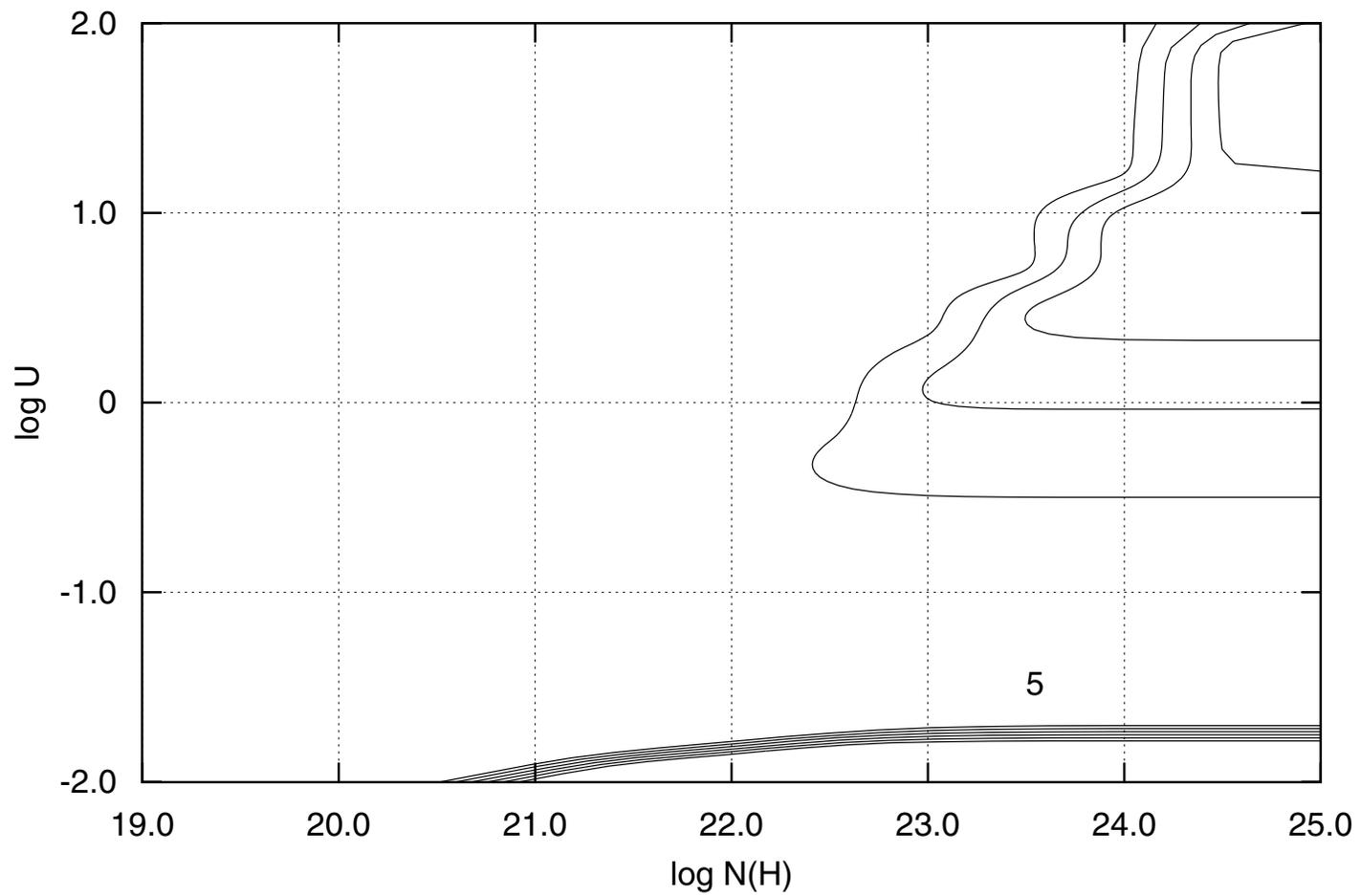

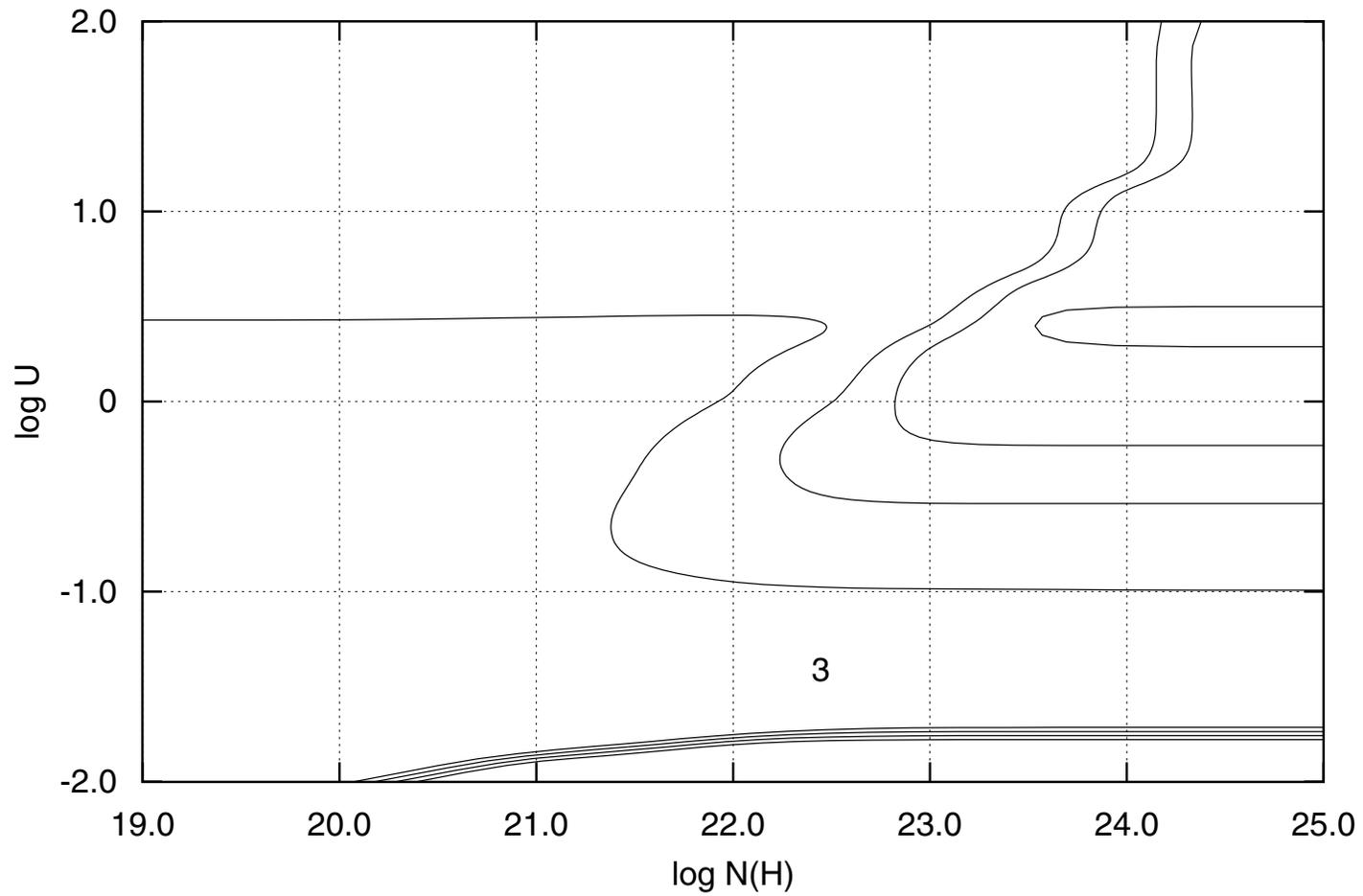

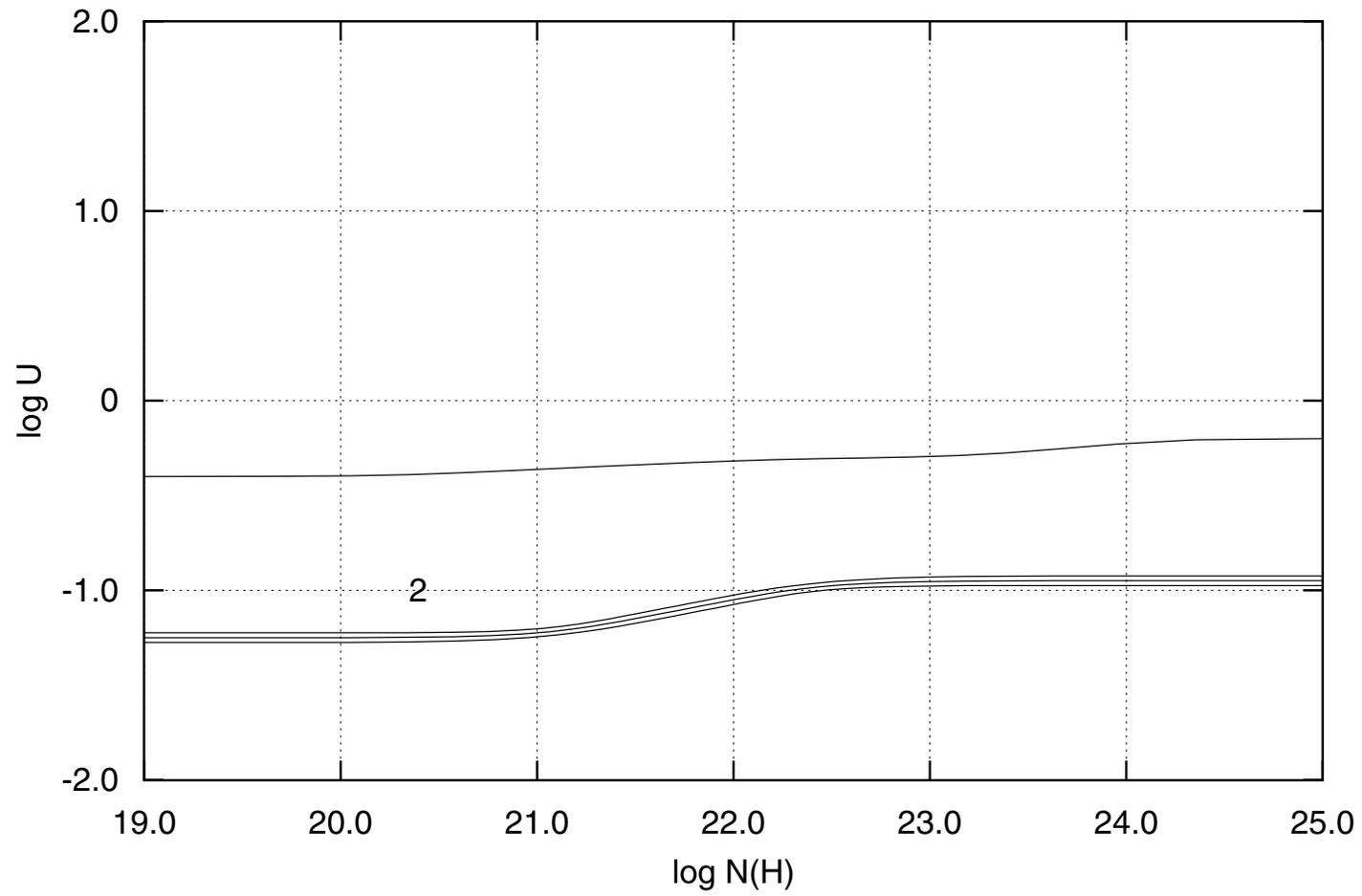

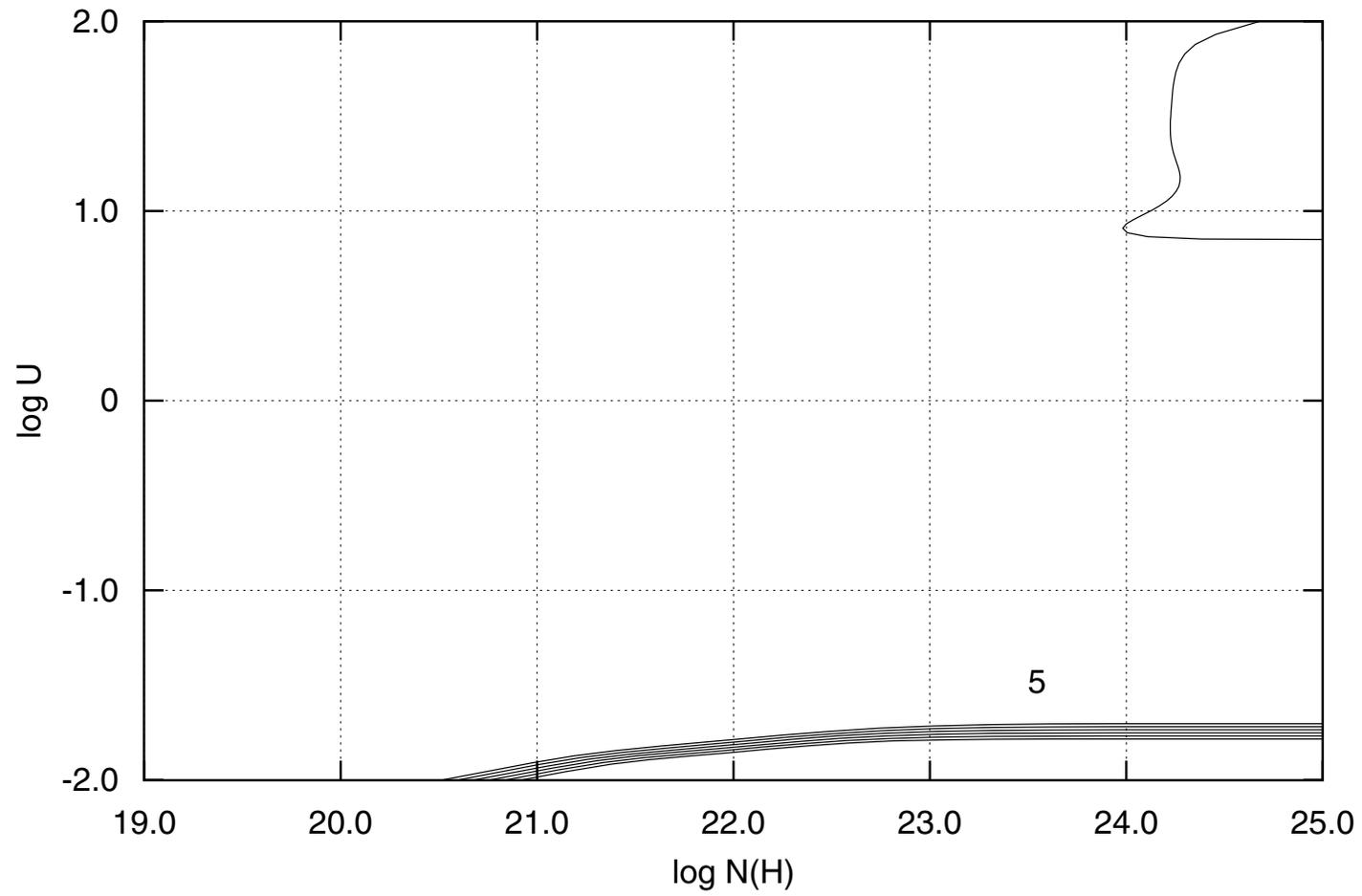

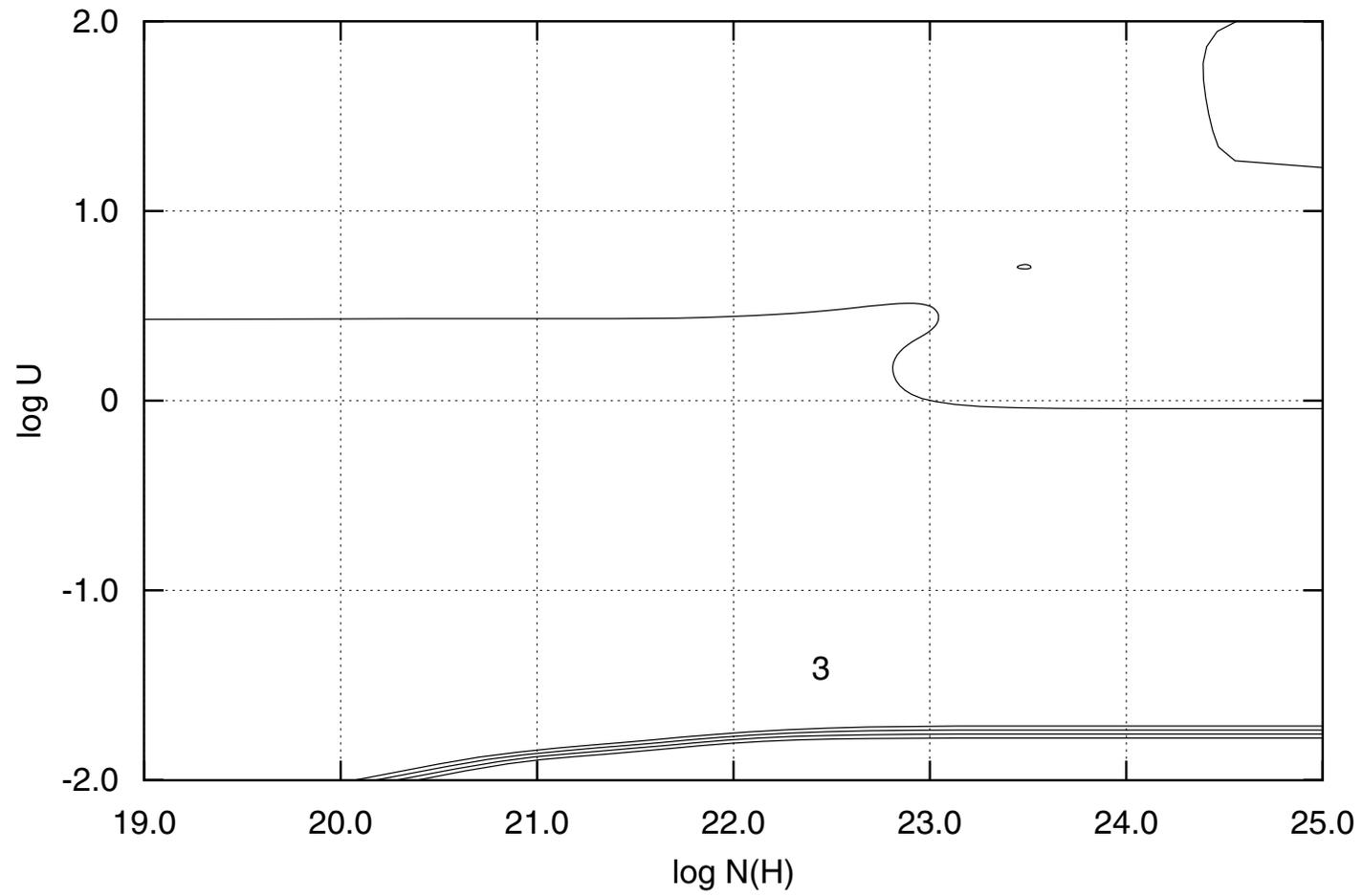

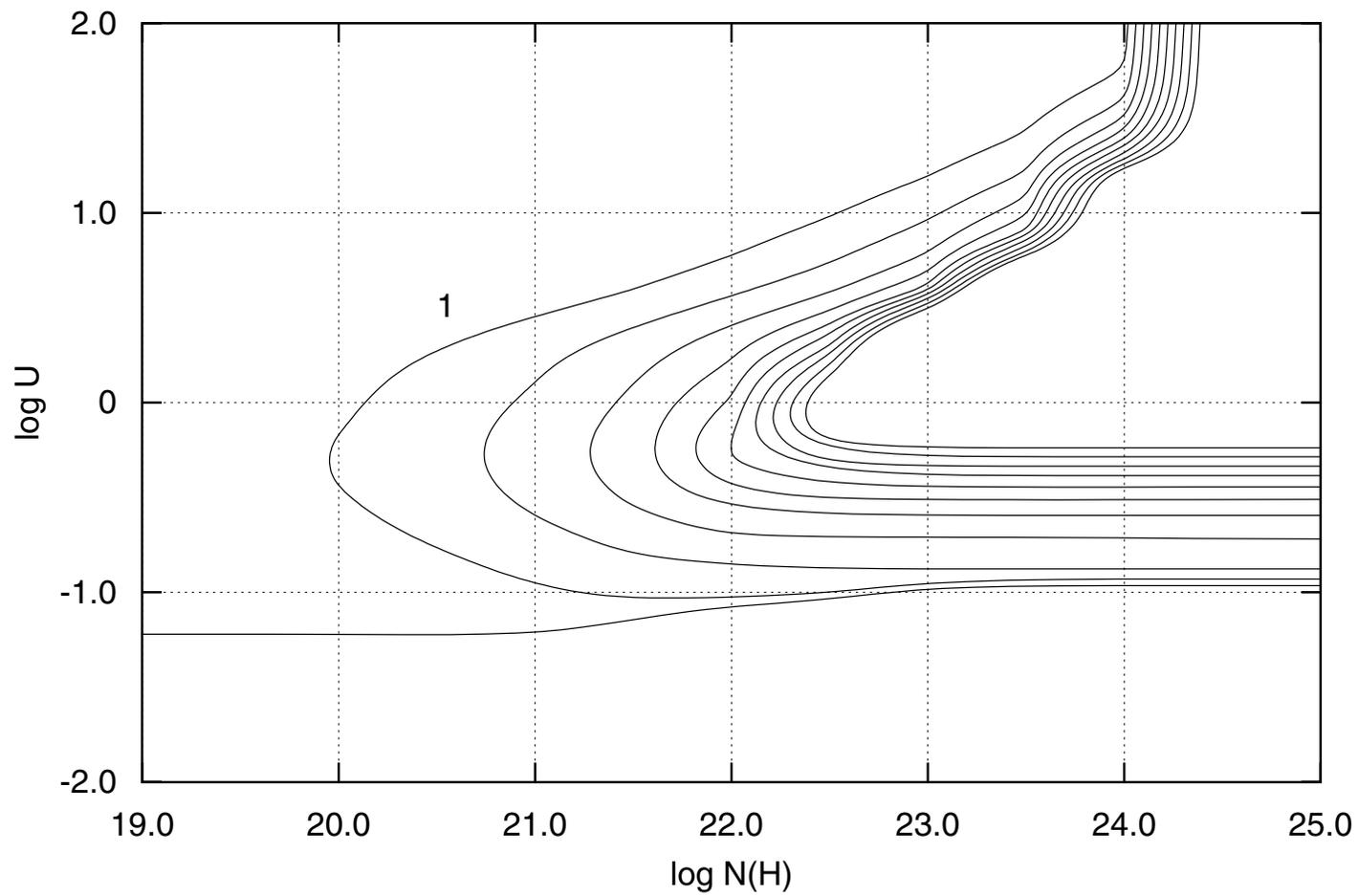

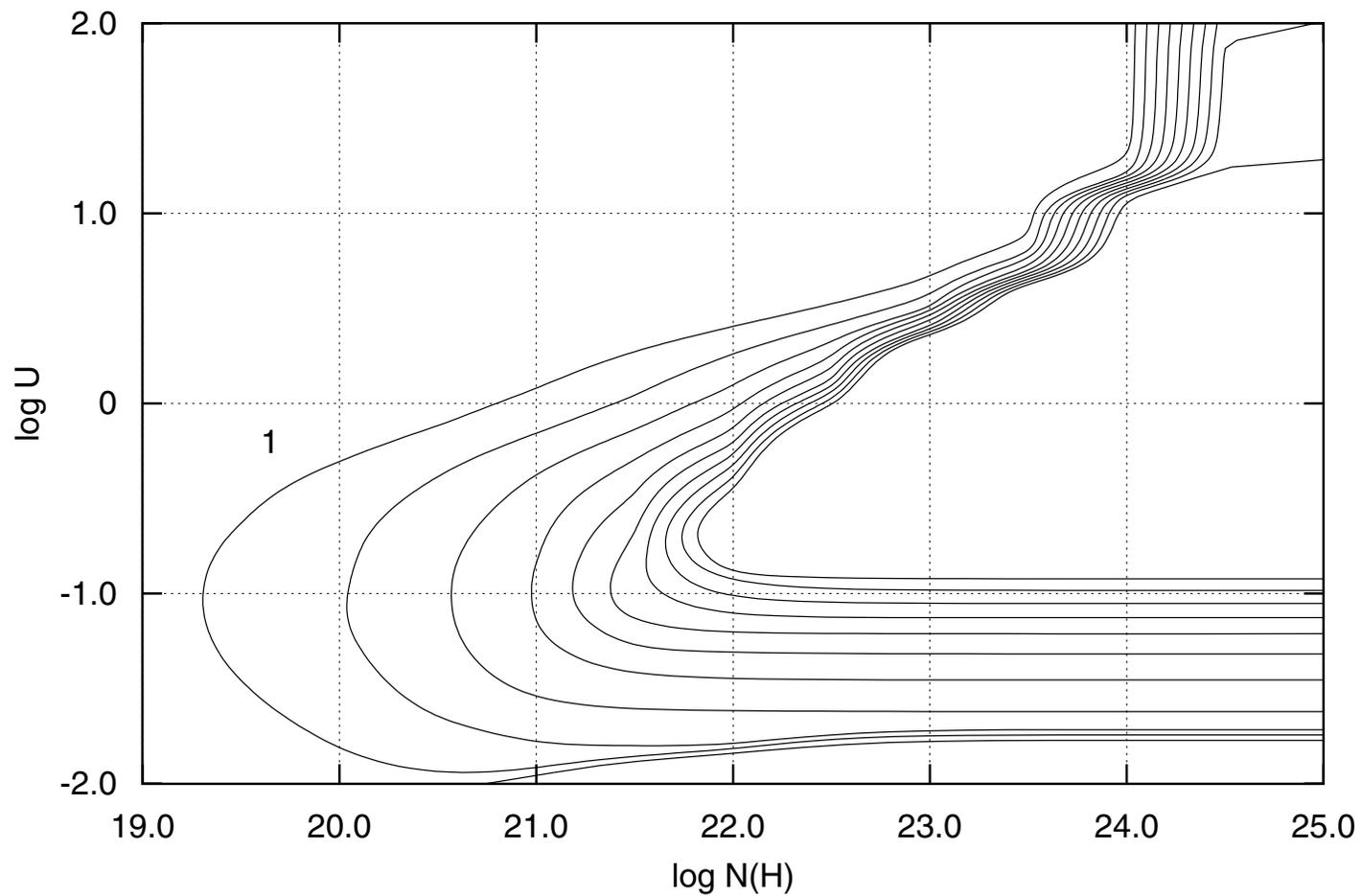

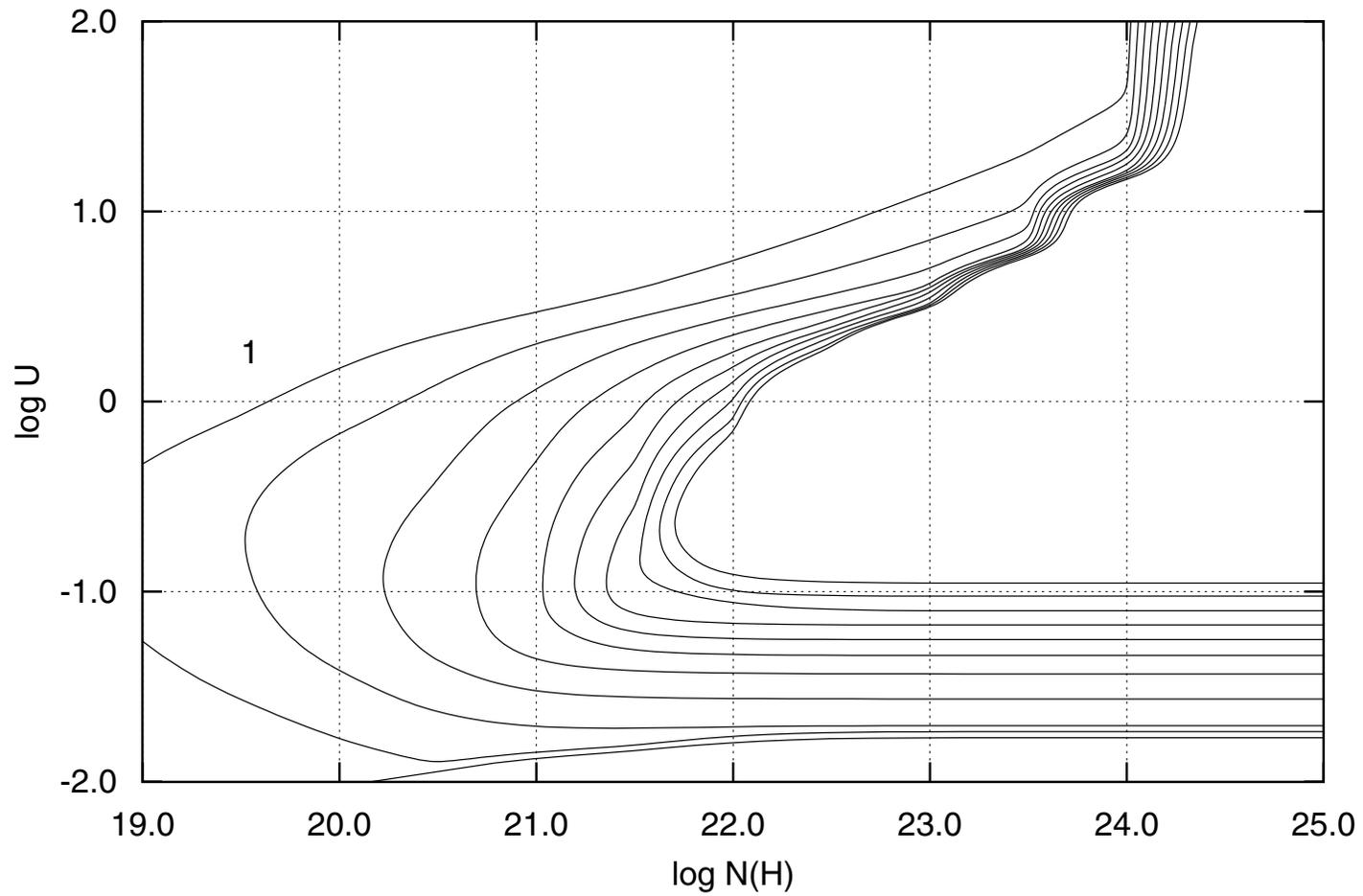

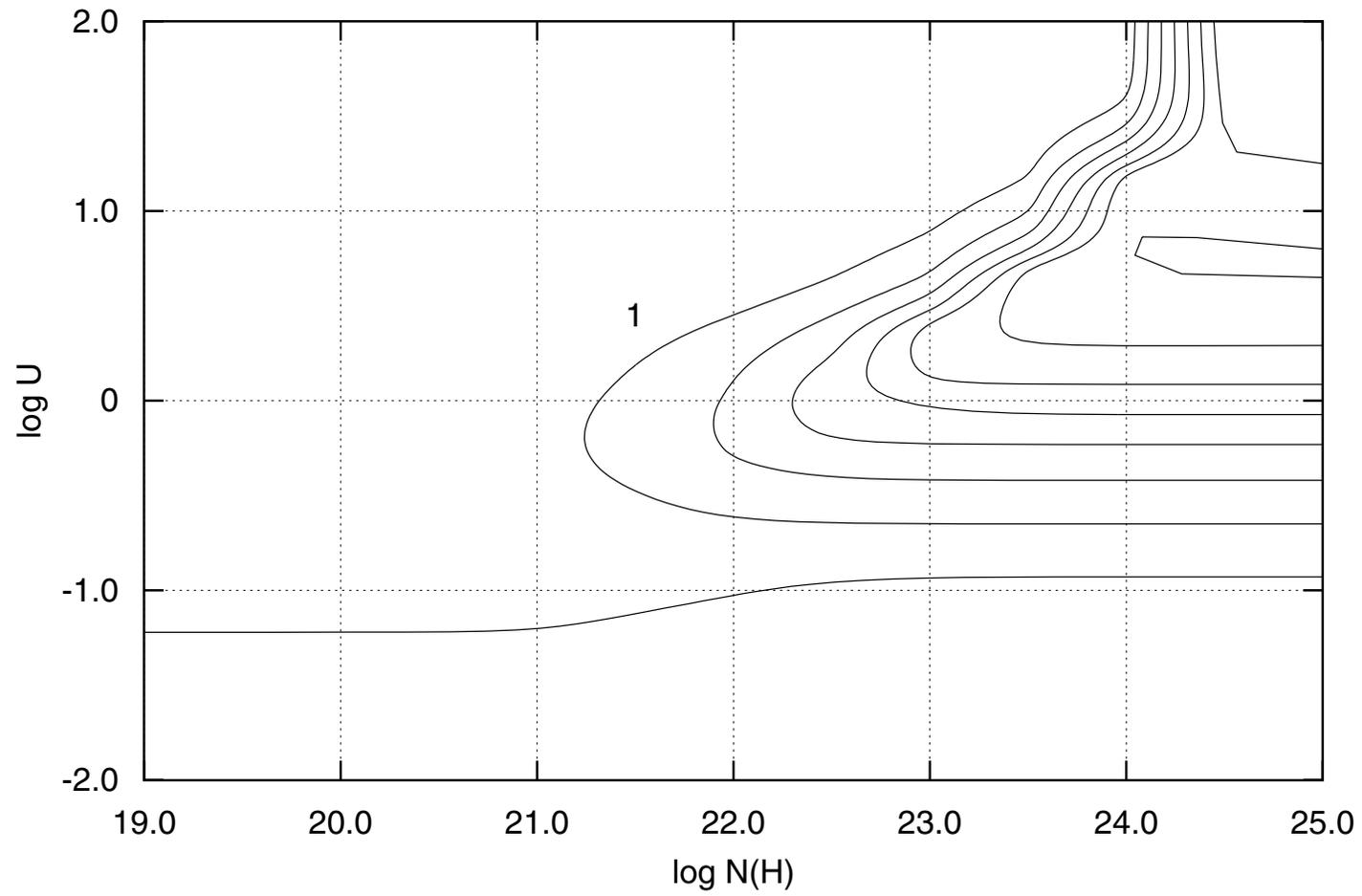

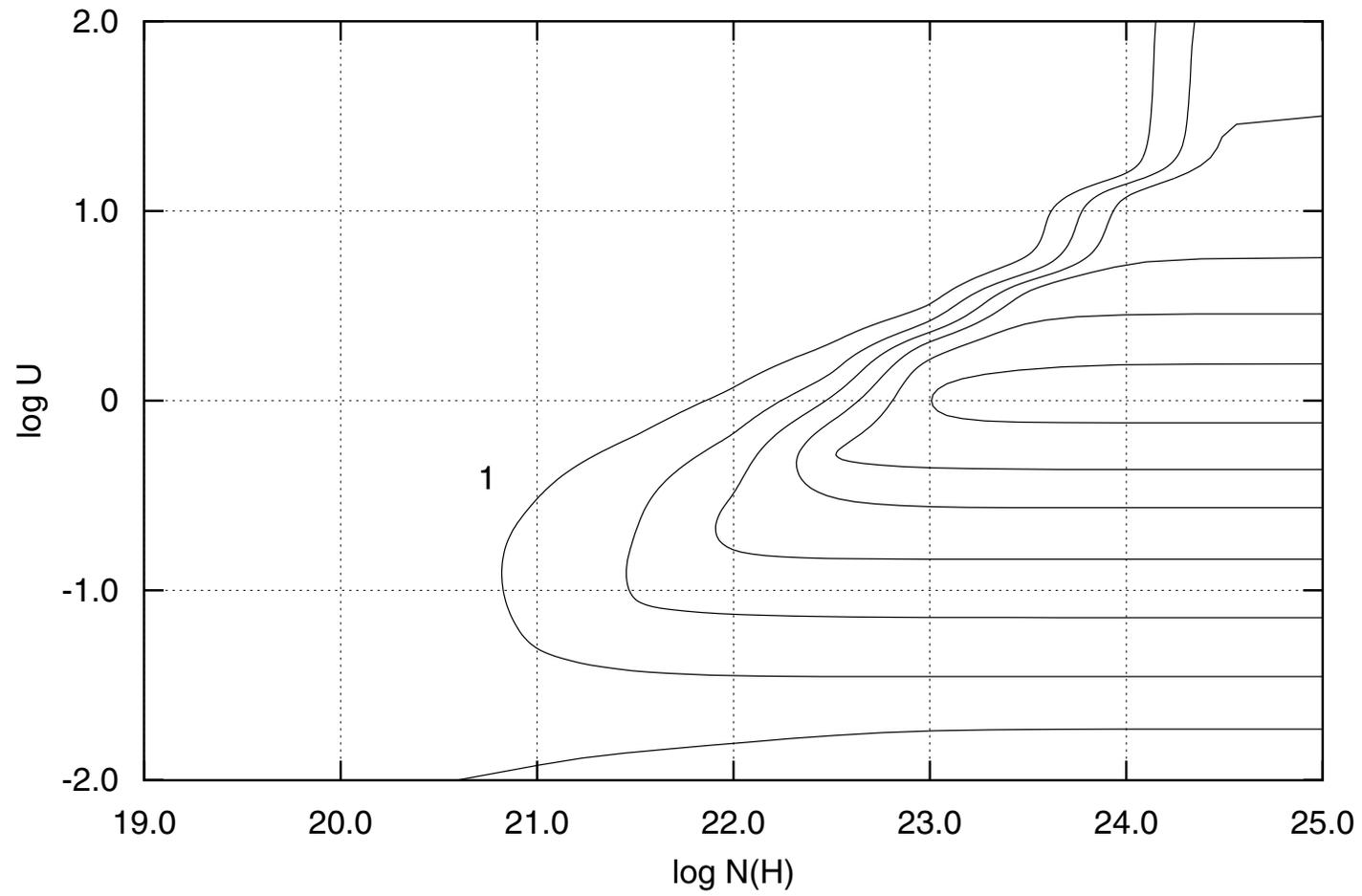

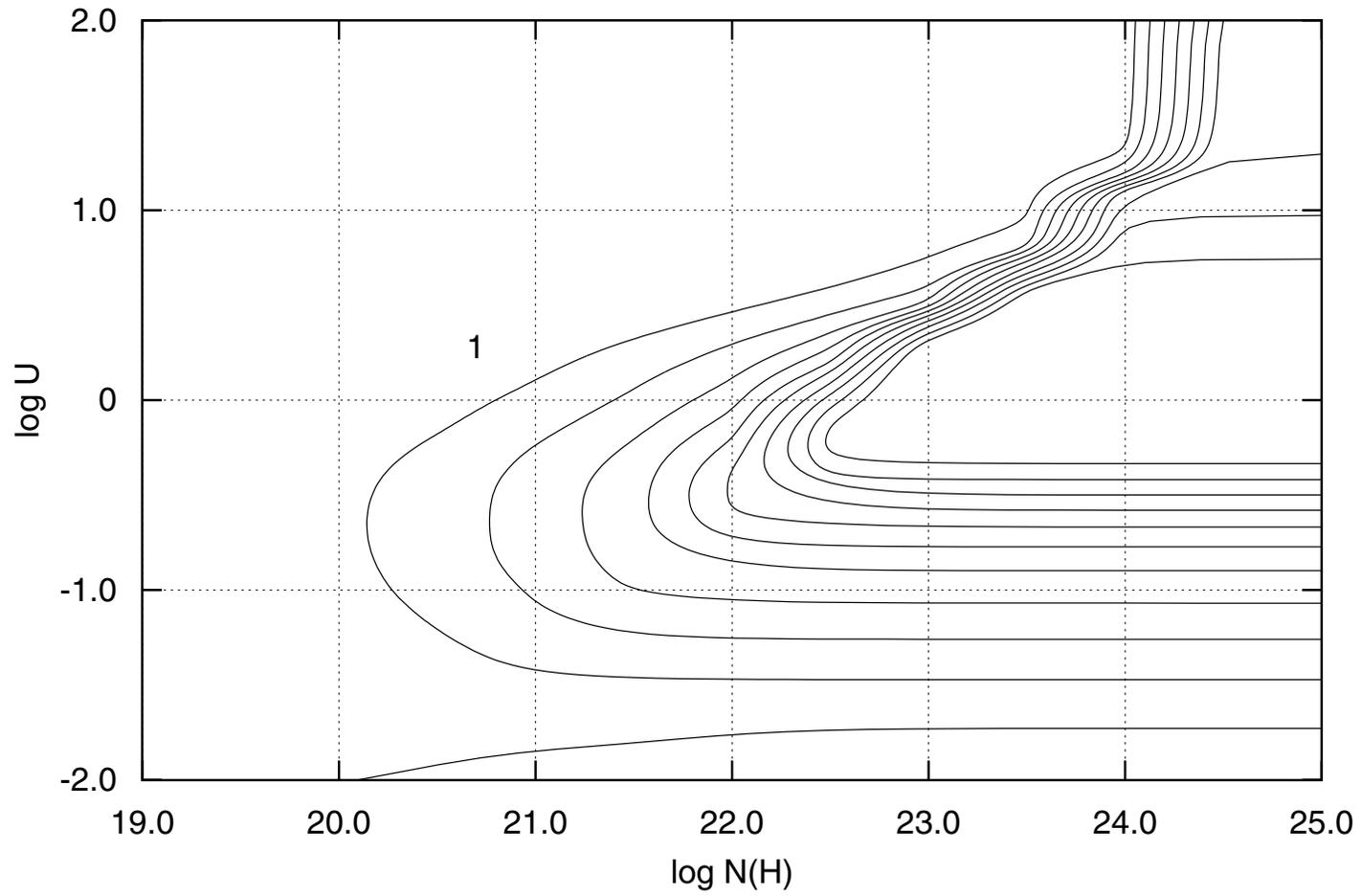

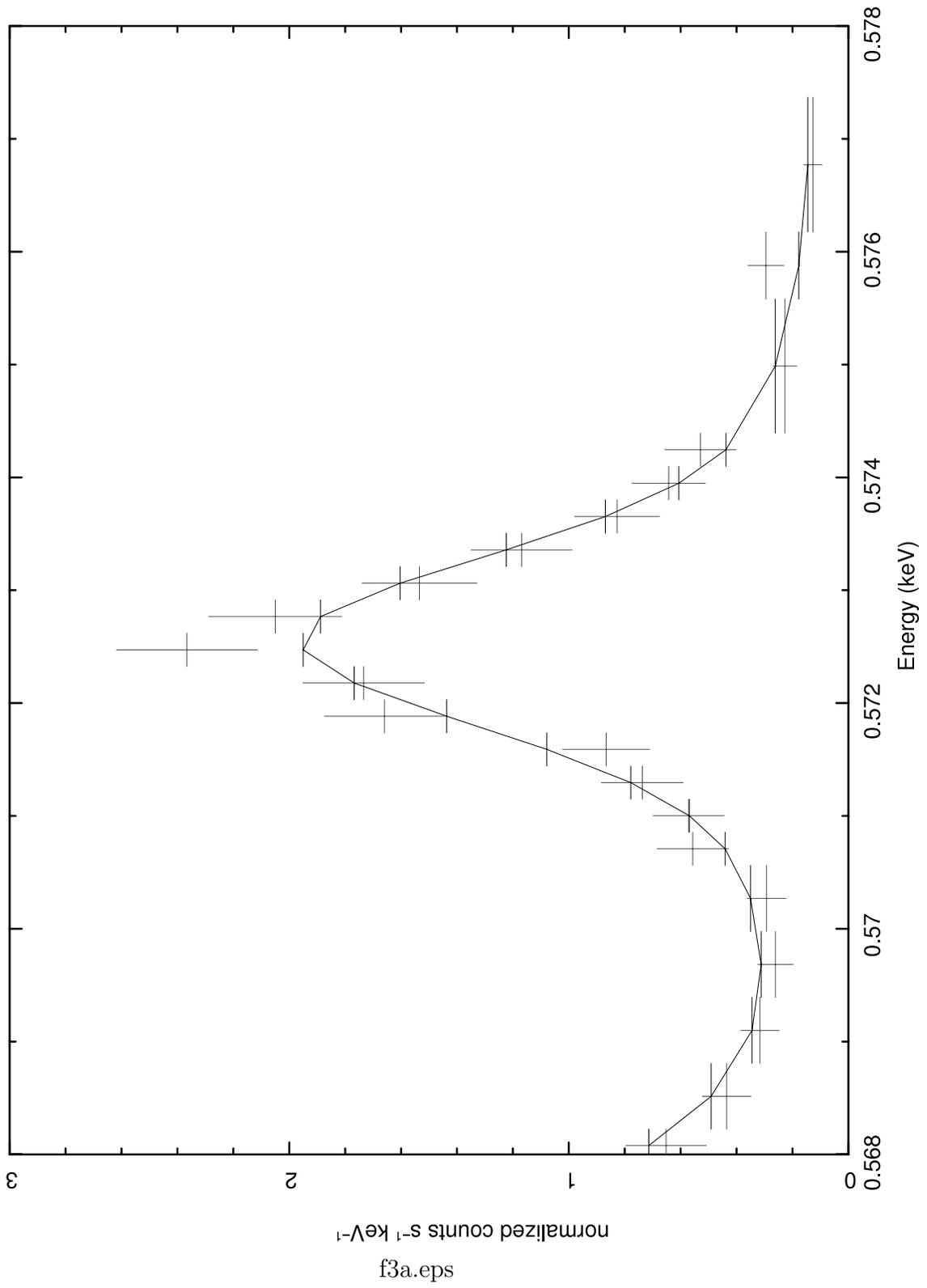

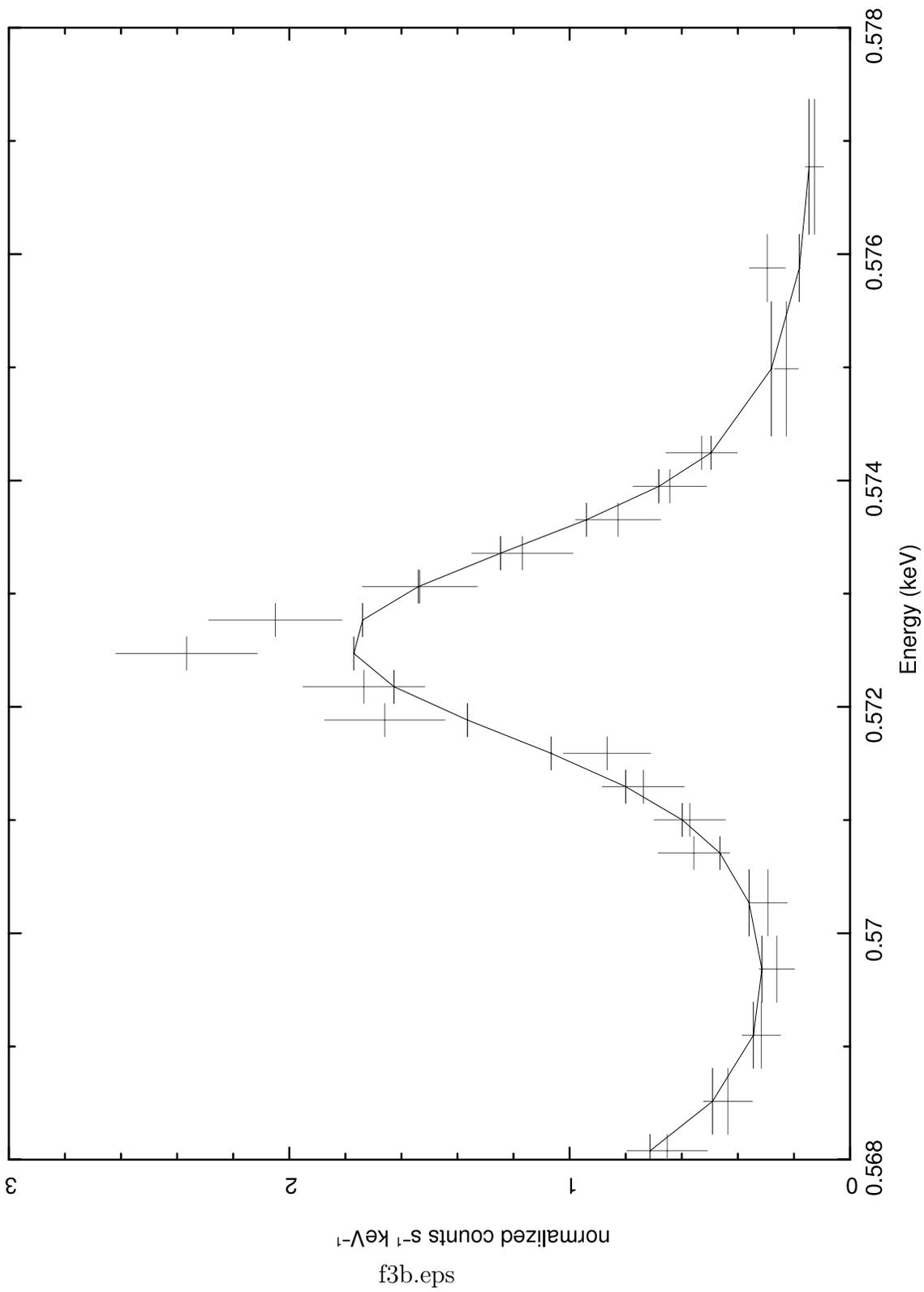

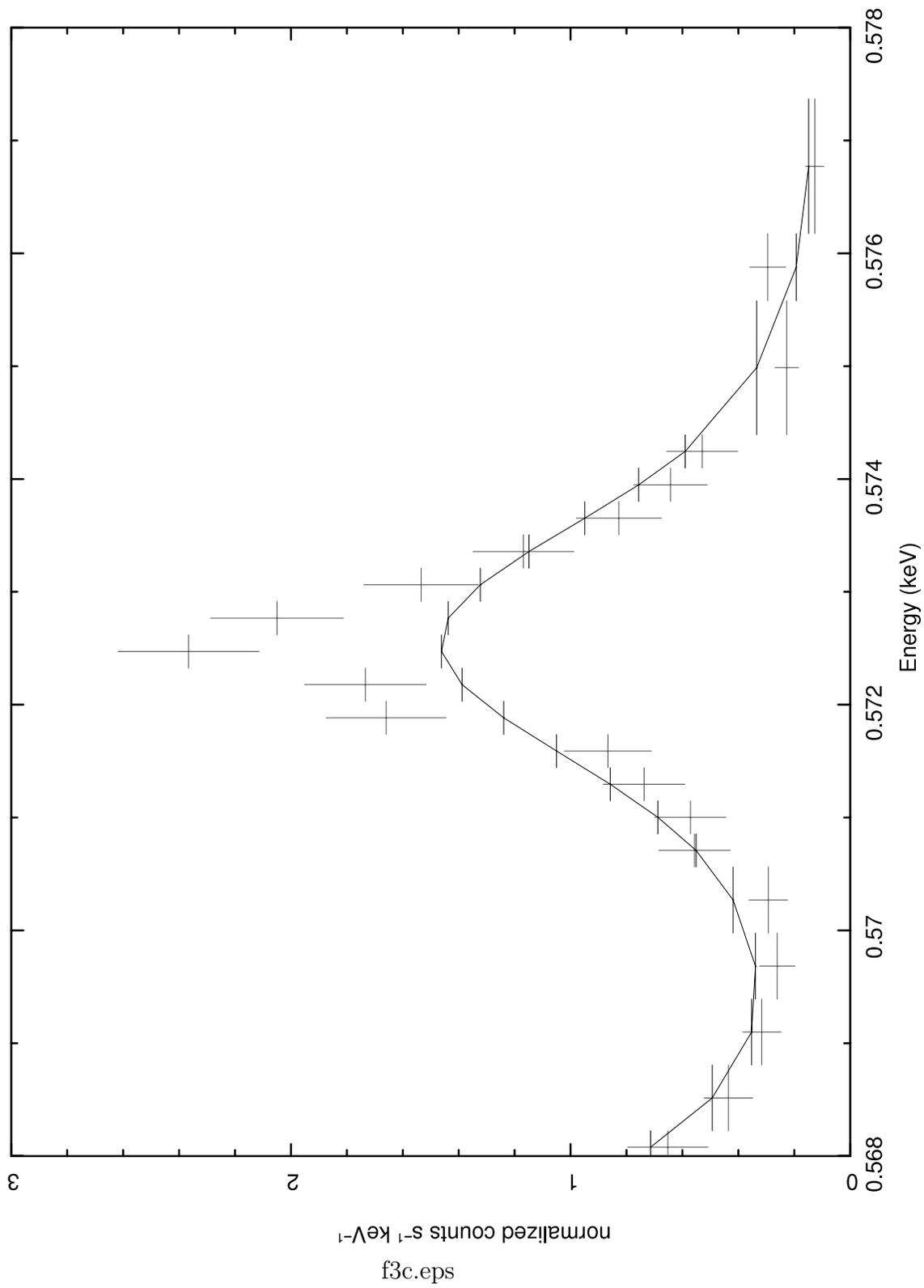

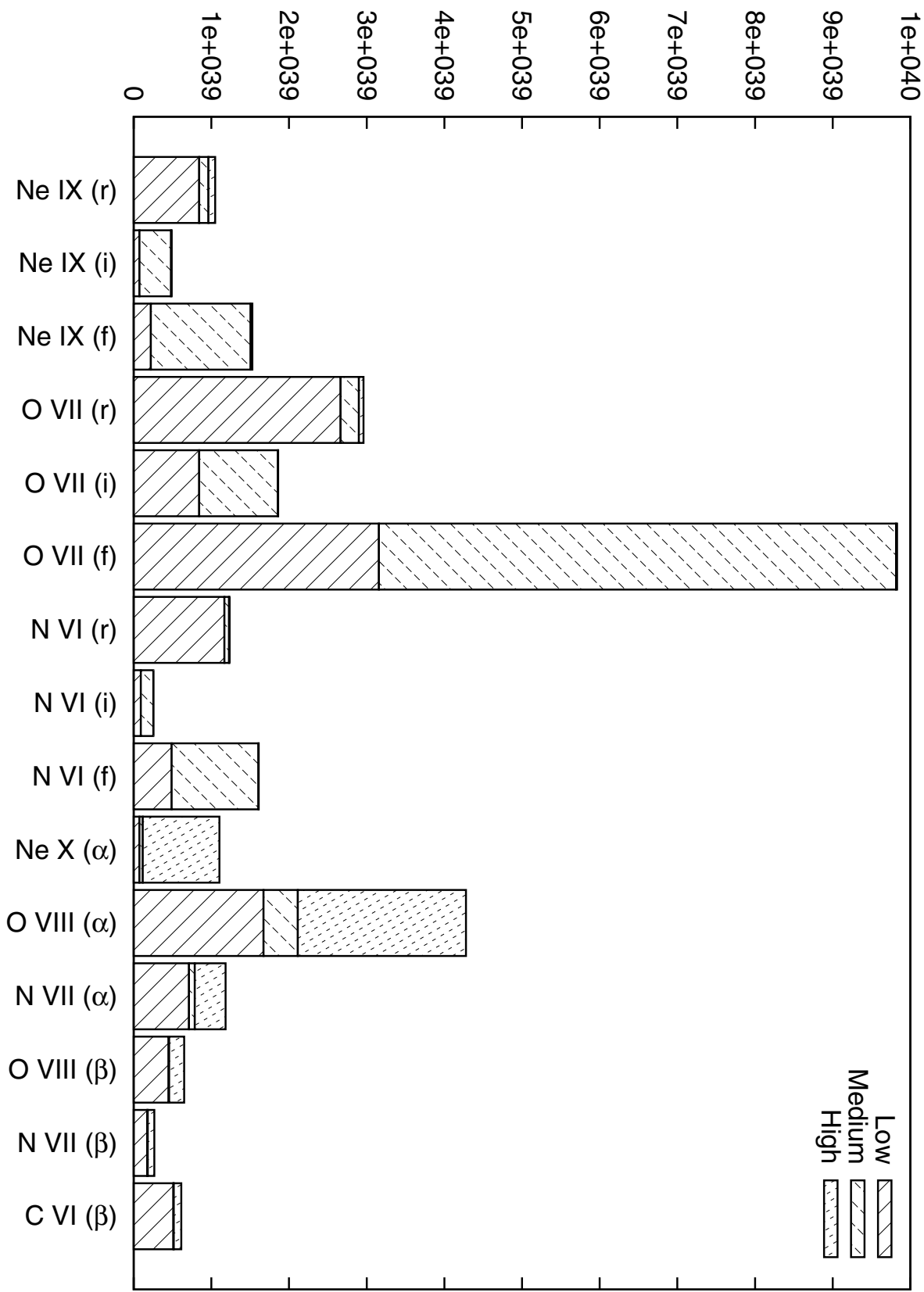

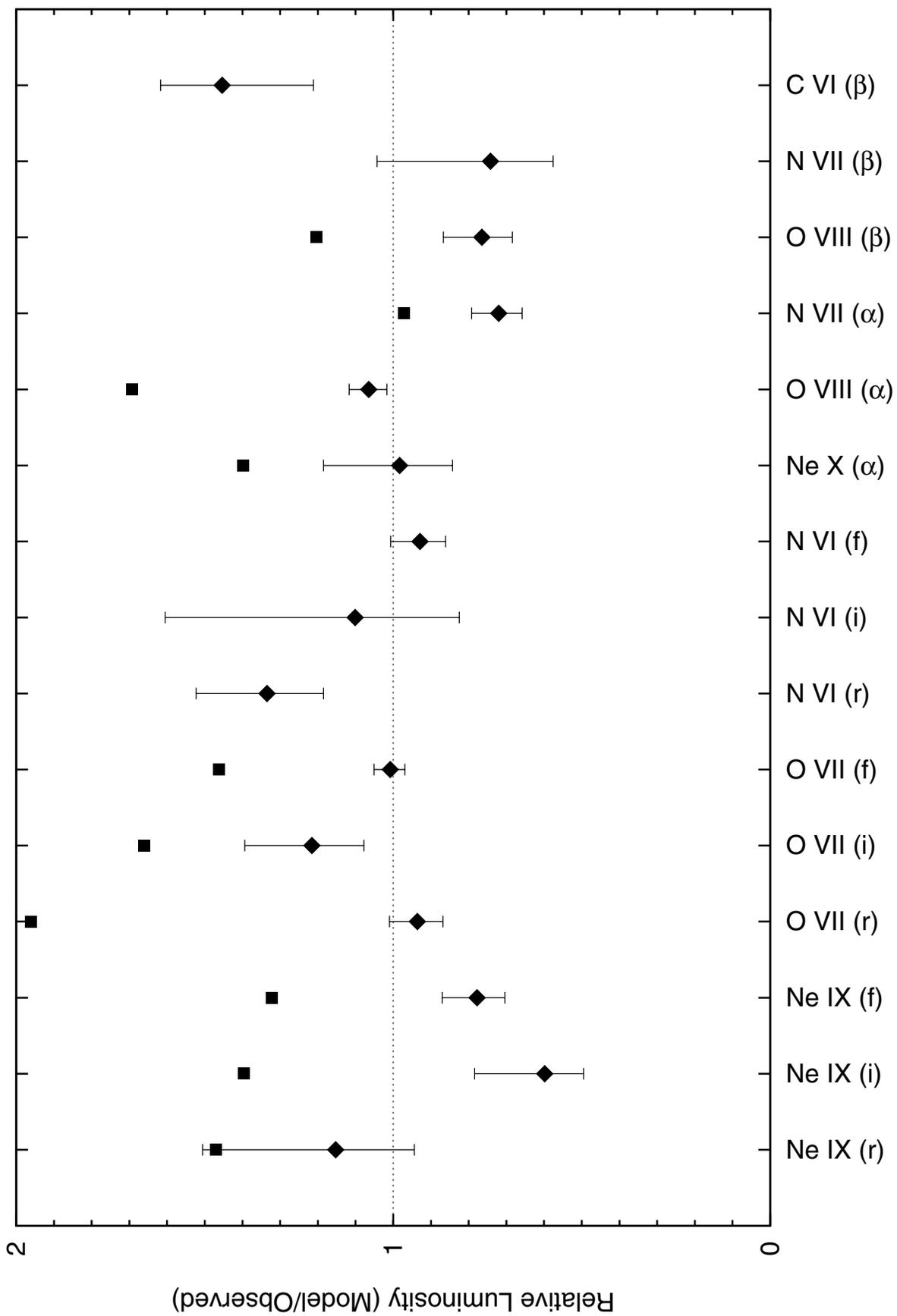